\documentclass[12pt]{article}
\usepackage[dvips]{graphicx}
\usepackage{pdproc}

  \makeatletter 
  \def\@cite#1{[#1]} 
  \makeatother    
  \newcommand{\lesssim}{ \mathop{}_{\textstyle \sim}^{\textstyle <} }
  
\begin{document}

\renewcommand{\thefootnote}{\alph{footnote}}

\title{
Correlated Isocurvature Fluctuation of Quintessence and 
Curvaton Scenario
}

\author{TAKEO MOROI$^1$ and TOMO TAKAHASHI$^2$}

\address{ 
$^1$Department of Physics, Tohoku University, 
Sendai 980-8578, Japan\\
$^2$Institute for Cosmic Ray Research,
University of Tokyo, Kashiwa 277-8582, Japan
}

\abstract{ We consider cosmic microwave background (CMB) anisotropy in
models with quintessence taking into account of isocurvature
fluctuation in the quintessence.  It is shown that, if the primordial
fluctuation of the quintessence has a correlation with the adiabatic
density fluctuations, CMB angular power spectrum $C_l$ at low
multipoles can be suppressed.  Possible scenario of generating
correlated mixture of the quintessence and adiabatic fluctuations is
also discussed.  }

\normalsize\baselineskip=15pt

\section{Introduction} Current cosmological observations suggest that
the present universe is dominated by an enigmatic component called
``dark energy."  Although many candidates for dark energy have been
proposed so far, a slowly evolving scalar field, dubbed as
``quintessence" is an interesting possibility. Since the quintessence
field is a scalar field, its amplitude may acquire primordial
fluctuation and isocurvature fluctuation may exist in the quintessence
sector, which affects the form of the CMB angular power spectrum.  In
this talk, we discuss the CMB angular power spectrum in models with
quintessence, paying particular attention to effects of the correlated
isocurvature fluctuation between quintessence and adiabatic
fluctuations \cite{Moroi:2003pq}.

\section{Framework}
Here we present the framework of our study. Although various
models of quintessence have been proposed, we adopt a simple
approximation for the quintessence potential of a quadratic form 
with a constant term\footnote{
Notice that this potential is a good
approximation for some quintessence models like the cosine-type one
\cite{JHEP9905-022}.
}
:
\begin{eqnarray}
V(Q) = V_0 + \frac{1}{2} m_Q^2 Q^2.
\end{eqnarray}
In our study, we consider $m_Q$ comparable to (or smaller than) the
present Hubble parameter.  In addition, $V_0$ is assumed to be of the
order of the present critical density or smaller.  With such a small
value of $m_Q$ (and $V_0$), slow-roll condition for $Q$ is satisfied
until very recently and energy fraction of the quintessence becomes
sizable only at the very recent epoch.  Since the quintessence is a
dynamical scalar field, its amplitude may fluctuate which can become a
new source of the cosmic density fluctuations and affects the CMB
angular power spectrum.

Evolution and effects of the primordial fluctuation of $Q$ have been
studied for the case where the primordial fluctuation of the
quintessence is not correlated with the adiabatic fluctuations
\cite{PRD64-083513,KMT}.  Fluctuation of the quintessence field can
be, however, correlated with the adiabatic fluctuations.  If some
correlation exists, effects of the primordial fluctuation of $Q$ on
the CMB angular power spectrum are expected to be different from those
in the uncorrelated case.  Hereafter, we discuss the effects of
quintessence fluctuation for the case where the correlation between
the fluctuation of $Q$ and the adiabatic fluctuations exists.

In order to parameterize the relative size of the primordial
quintessence fluctuation and the adiabatic fluctuations, we
define\footnote{ 
Strictly speaking, the following expression is
valid only for the case where $\delta Q_{\rm init}$ and $\Psi_{\rm
RD}$ are fully correlated.  For the case where $\delta Q_{\rm init}$
and $\Psi_{\rm RD}$ are uncorrelated, for example, it should be
understood as 
$r_Q=\sqrt{\langle\delta Q_{\rm init}^2\rangle}/M_*
\sqrt{\langle\Psi_{\rm RD}^2\rangle}$.  
} 
\begin{eqnarray} r_Q
\equiv \frac{\delta Q_{\rm init}}{M_* \Psi_{\rm RD}}.  
\end{eqnarray}

\begin{figure}[htb]
\begin{center}
\includegraphics*[width=10cm]{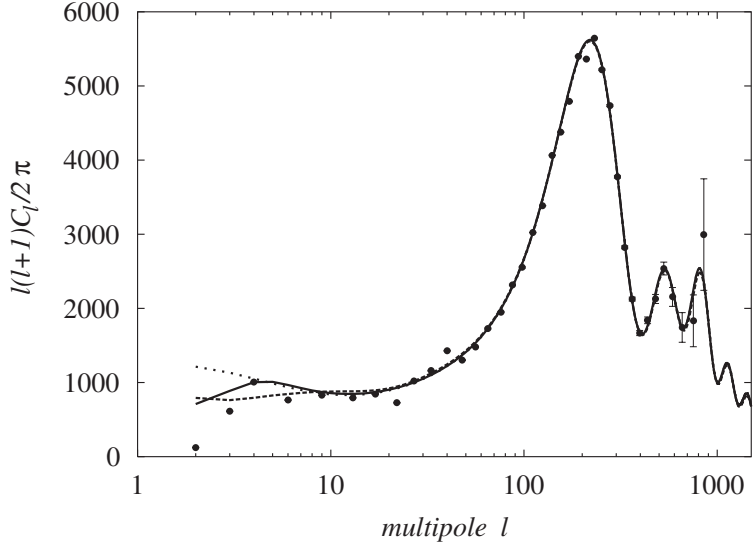}
\caption{The CMB angular power spectrum for  the correlated
mixture of the quintessence and adiabatic fluctuations.  We take (a)
$m_Q=10^{-42}\ {\rm GeV}$, $V_0=0$ and $r_Q=400$ with
$\Omega_b=0.046$, $\Omega_m=0.27$, $h=0.72$, and $\tau=0.166$ (solid), 
and (b) $m_Q=10^{-40}\ {\rm GeV}$,
$V_0=3.0\times 10^{-47}\ {\rm GeV}$ and $r_Q=2.5$ with
$\Omega_b=0.048$, $\Omega_m=0.27$, $h=0.7$, and $\tau=0.2$ (dashed).
The full correlation between $\delta Q_{\rm init}$ and $\Psi_{\rm RD}$
is assumed.  Result for the $\Lambda$CDM model is
also shown in the dotted line.   For comparison, we also plot the data points
measured by the WMAP \protect\cite{wmap}.  (The errors are
measurement errors only.)}
\label{fig:Cl}
\end{center}
\end{figure}

Here, $\delta Q_{\rm init}$ is the primordial fluctuation of $Q$,
$\Psi$ denotes the fluctuation of the $(0,0)$ component of the metric
in the Newtonian gauge: $g_{00}=a^2(1+2\Psi)$ with $a$ being the scale
factor, and $M_*$ is the reduced Planck scale.  In addition,
$\Psi_{\rm RD}$ is the metric perturbation related to the adiabatic
density fluctuation in the radiation-dominated epoch.  We assume that
$\Psi_{\rm RD}$ is (almost) scale-invariant.  If we calculate the CMB
angular power spectrum with non-vanishing values of $\delta Q_{\rm
init}$ and $\Psi_{\rm RD}$, we obtain \begin{eqnarray} C_l = C_l^{\rm
(adi)} + C_l^{\rm (corr)} + C_l^{\rm (uncorr)}.  \end{eqnarray} Here,
$C_l^{\rm (adi)}$ is the result with purely adiabatic density
fluctuations.  $C_l^{\rm (uncorr)}$ is the CMB angular power spectrum
purely generated from $\delta Q_{\rm init}$, while $C_l^{\rm (corr)}$
parameterizes the effects of correlation.  When $\delta Q_{\rm init}$
and $\Psi_{\rm RD}$ are uncorrelated, $C_l^{\rm (corr)}=0$.
Furthermore, $C_l^{\rm (uncorr)}$ is increased at low multipoles.  As
a result, the total CMB angular power spectrum may be significantly
enhanced at low multipoles when no correlation is assumed \cite{KMT}.
If $\delta Q_{\rm init}$ and $\Psi_{\rm RD}$ have correlation,
$C_l^{\rm (corr)}$ plays important roles.  In Fig.\ \ref{fig:Cl}, we
show the CMB angular power spectrum for the case where $\delta Q_{\rm
init}$ and $\Psi_{\rm RD}$ are fully correlated.  As one can see, in
this case, sizable suppression of the low multipoles is possible
compared to the $\Lambda$CDM model.  Notice that, in the uncorrelated
case, such a suppression of $C_l$ at the low multipoles cannot be
realized.

\section{A possible model}
Here we present a possible scenario which generates the correlated
fluctuations.  We use the fact that fluctuations of two scalar fields
can be correlated if they have a mixing during the inflation.  We
consider the case with two scalar fields, $Q$ and $\phi$.  $Q$ and
$\phi$ are defined as mass eigenstates in the present universe.  In
the early universe (i.e., for example, during inflation), however,
Hubble-induced interaction may cause a mixing between $Q$ and $\phi$
and hence the mass eigenstates may be linear combinations of them.  We
denote the mass eigenstates as $\xi$ and $\eta$, and define the mixing
angle $\theta$ as \begin{eqnarray} \left( \begin{array}{c} \xi \\ \eta
\end{array} \right) = \left( \begin{array}{cc} \cos\theta (t) &
-\sin\theta (t) \\ \sin\theta (t) & \cos\theta (t) \end{array} \right)
\left( \begin{array}{c} Q \\ \phi \end{array} \right).  \end{eqnarray}
In our model, $\theta (t)$ varies from non-vanishing value during
inflation $\theta_{\rm inf}$ to the present value $\theta_{\rm
now}=0$.  If the mass of $\eta$ is large during inflation while that
of $\xi$ is negligible, only the $\xi$ field acquires the quantum
fluctuation as $ \delta \xi_{\rm inf} =H_{\rm inf}/2\pi, \delta
\eta_{\rm inf} = 0, $ where $H_{\rm inf}$ is the expansion rate during
the inflation.  Assuming that $\theta$ rapidly changes from
$\theta_{\rm inf}$ to $0$ at the end of inflation, primordial
fluctuation of $Q$ and $\phi$ are given by $ \delta Q_{\rm init} =
\delta \xi_{\rm inf}\cos\theta_{\rm inf}, \delta \phi_{\rm init} =
\delta \xi_{\rm inf} \sin\theta_{\rm inf}, $ then correlated
fluctuations are generated in $Q$ and $\phi$.

The above situation may be realized if the potential of the scalar
fields is of the form
\begin{eqnarray}
V = V_0 + \frac{1}{2} m_Q^2 Q^2 + \frac{1}{2} m_\phi^2 \phi^2 
+ V_{\rm Hubble},
\label{V_tot}
\end{eqnarray}
where
$
V_{\rm Hubble} = H_{\rm vac}^2 
( Q \sin \theta_{\rm inf} + \phi \cos \theta_{\rm inf} )^2.
$
Here, $V_{\rm Hubble}$ is the Hubble-induced interaction which is
effective only during the inflation with $H_{\rm vac}$ being the
Hubble parameter induced by the ``vacuum energy;'' $H_{\rm vac}=H_{\rm
inf}$ and $H_{\rm vac}\simeq 0$ for during and after the inflation,
respectively.  With this potential, one of the mass eigenstates
$\eta\simeq\phi+\theta_{\rm inf}Q$ acquires an effective mass
comparable to the expansion rate and its quantum fluctuation during
inflation becomes negligibly small.  Other mass eigenstate $\xi\simeq
Q-\theta_{\rm inf}\phi$, on the contrary, stays almost massless and it
acquires the quantum fluctuation.

If the decay rate of the inflaton field is larger than $m_\phi$, slow
roll condition is satisfied for $\phi$ at the time of the inflaton
decay.  In this case, $\delta\phi_{\rm init}$ may become the dominant
source of the adiabatic fluctuations.  In order to generate the
adiabatic fluctuations from the fluctuation of $\phi$, we can use the
curvaton mechanism \cite{curvaton} where the primordial fluctuation of
the curvaton becomes the dominant source of the adiabatic
fluctuations\footnote{
Here, we do
not identify $\phi$ as inflaton; in our model, $\phi$ should acquire
large effective mass during inflation and hence $\phi$ cannot be the
inflaton.
}
. Indeed, if the energy density of $\phi$ once dominates the universe,
$\phi$ plays the role of curvaton and the metric perturbation in the
radiation dominated epoch is given by $ \Psi_{\rm
RD} = -(4/9) (\delta\phi_{\rm init}/\phi_{\rm init}) $
where $\phi_{\rm init}$ is the initial amplitude of $\phi$ determined
during the inflation \cite{Psi(curv)}.  As a result, correlated
mixture of adiabatic and quintessence fluctuations is generated.  In
this model, the $r_Q$ parameter is estimated as 
\begin{eqnarray} r_Q =
\frac{1}{2\pi} \frac{H_{\rm inf}}{M_* \Psi_{\rm RD}} 
\cos\theta_{\rm inf}.  
\end{eqnarray} 

Notice that, using $\Psi_{\rm RD}\sim O(10^{-5})$ and the upper bound
$H_{\rm inf}/M_*\lesssim 7\times 10^{-5}$ \cite{H_max}, $r_Q\lesssim
1$ in this simple model.  Larger value of $r_Q$ is, however, possible
if we extend the model.  For example, if the coefficient of the
kinetic term of $Q$ varies after the inflation, value of $\delta Q$
(for the canonically normalized field) also changes \cite{Moroi:2003pq}.

\section{Summary}

We have seen that the CMB angular power spectrum at small $l$ can be
suppressed if the primordial fluctuation of the quintessence has
correlation with the adiabatic density fluctuations\footnote{
For subsequent work along this line, see Ref.\ \cite{Gordon:2004ez}.
}
.  We have also
pointed out that such a correlation may be generated during inflation
if the quintessence field has some mixing with other scalar field
which is responsible for generating the adiabatic density
fluctuations.

\bigskip

\noindent
{\bf Acknowledgments:}
T.T. would like to thank the Japan Society for Promotion of Science
for financial support. This work was partially supported by the
Grand-in-Aid of the Ministry of Education, Science, Sports and Culture
of Japan No.~15540247 (T.M.).


\begin{thebibliography}{99}

\bibitem{Moroi:2003pq}
T.~Moroi and T.~Takahashi,
Phys.\ Rev.\ Lett.\  {\bf 92}, 091301 (2004)

\bibitem{JHEP9905-022}
    J.~E.~Kim,
    JHEP {\bf 9905} (1999) 022.

\bibitem{PRD64-083513}
    L.~R.~Abramo and F.~Finelli,
    Phys.\ Rev.\ D {\bf 64} (2001) 083513.

\bibitem{KMT}
    M.~Kawasaki, T.~Moroi and T.~Takahashi,
    Phys.\ Rev.\ D {\bf 64} (2001) 083009; 
    Phys.\ Lett.\ B {\bf 533} (2002) 294.

\bibitem{wmap}
    G.~Hinshaw {\it et al.},
    Astrophys.\ J.\ Suppl.\  {\bf 148}, 135 (2003).

\bibitem{curvaton}
    K.~Enqvist and M.~S.~Sloth,
    Nucl.\ Phys.\ B {\bf 626} (2002) 395;
    D.~H.~Lyth and D.~Wands,
    Phys.\ Lett.\ B {\bf 524} (2002) 5;
    T.~Moroi and T.~Takahashi,
    Phys.\ Lett.\ B {\bf 522} (2001) 215
    [Erratum-ibid.\ B {\bf 539} (2002) 303].

    
\bibitem{Psi(curv)}
    T.~Moroi and T.~Takahashi, in Ref.\ \cite{curvaton};
    Phys.\ Rev.\ D {\bf 66} (2002) 063501.


\bibitem{H_max}
    H.~V.~Peiris {\it et al.},
    Astrophys.\ J.\ Suppl.\  {\bf 148} (2003) 213;
    V.~Barger, H.~S.~Lee and D.~Marfatia,
    Phys.\ Lett.\ B {\bf 565} (2003) 33;
    S.~M.~Leach and A.~R.~Liddle,
    Phys.\ Rev.\ D {\bf 68}, 123508 (2003)

\bibitem{Gordon:2004ez}
C.~Gordon and W.~Hu,
arXiv:astro-ph/0406496.

\end{thebibliography}
\end{document}